\documentclass[12pt]{article} % 10pt is ignored!

\usepackage{epsfig}
\usepackage{graphicx}% Include figure files
\usepackage{amssymb}

\begin{document}

\title{A WDM model for the evolution of galactic halos}

\author{L. Acedo\thanks{E-mail: luiacrod@imm.upv.es}\\
Instituto de Matem\'atica Multidisciplinar,\\ 
Universidad Polit\'ecnica de Valencia,\\ Camino de Vera s/n, 46022 Valencia, Spain\\
}

\maketitle

\begin{abstract} It is a well-known fact that the gravitational 
effect of dark matter in galaxies is only noticeable when the orbital
accelerations drop below $a_0 \simeq 2\times 10^{-8}$ cm s$^{-1}$ (Milgrom's Law). This 
peculiarity of the dynamic behaviour of galaxies was initially ascribed to a modification
of Newtonian dynamics (MOND theory) and, consequently, it was used as an argument to 
criticize the dark matter hypothesis. In our model, warm dark matter is composed
by collisionless Vlasov particles with a primordial typical velocity
$\simeq 330$ km s$^{-1}$ and, consequently, they evaporated
from galactic cores and reorganized in halos with a cusp at a finite distance from the galactic center (in contrast with Cold Dark Matter
simulations which predict a cusp at the center of galaxies). This is confirmed by mean-field N-body simulations
of the self-gravitating Vlasov dark matter particles in the potential well of the baryonic core. The 
rest mass of these particles, $\mu$, is determined from a kinetic theory of the early universe with a cosmological constant. We find that
$\mu$ is in the range of a few keV. This result makes sterile neutrinos the best suited candidates for
the main component of dark matter.
\end{abstract}

{\bf Keywords:} dark matter simulations, galaxy evolution, physics of the early universe

%\date{November 4, 2011}

\section{Introduction}
\label{Sect_I}
In the pioneer work of Zwicky 
\cite{zwicky} he presented the results of
the first detailed observations of the dynamics of the Coma galaxy cluster.
In this work, it was hypothesized that most of the matter in these clusters
must be dark because luminous matter is not sufficient to account for the
orbital velocities as deduced from Newtonian dynamics. Zwicky's dark matter
hypothesis was consigned to oblivion by the astrophysical community until the
end of the seventies and the beginning of the eighties of the past century.
At that time, a series of exhaustive measurements of the Doppler shift for the
$21$ cm hydrogen line in the galactic gas clouds for the Milky Way and other galaxies
proved, beyond any doubt, that the rotation curves of galaxies were
anomalous \cite{faber,rubin,caldwell}. This anomaly is characterized by a 
flat asymptotic region in the rotation curve corresponding to a constant orbital
velocity of stars and gas clouds, $V_\infty$. This is the case even for distances of
thousands of kpcs beyond the luminous core of the galaxy. If galaxies were composed
only of the observed luminous matter we will expect a decrease of the orbital
velocity $\propto r^{-1/2}$, $r$ being the distance to the galactic centre. Consequently, the
most natural explanation of this phenomenon is the presence of Zwicky's dark matter, inferred
from the observation of clusters of galaxies, also inside individual galaxies in the form of disperse
halos with a volumetric density decreasing as $r^{-2}$.

Just two years after the acceptance of this evidence by the astrophysical community, Milgrom proposed
an alternative explanation for the anomalous rotation curves of galaxies based upon a modification
of Newtonian dynamics for small accelerations (MOND theory) \cite{milgroma}. A careful
analysis of the rotation curves published by Faber \& Gallagher \cite{faber} revealed that the
effect of the supposed dark matter seems to activate only when the orbital acceleration is smaller than
$a_0 \approx 2 \times 10^{-8}$ cm s$^{-2}$ (do not confuse with the curvature radius of the Universe usually 
denoted by the same symbol). Consequently, Milgrom modified Newton's second law of dynamics 
by assuming that the gravitational mass of an object is a function $m(a/a_0)$. From his theory, Milgrom also
derived the following equation
\begin{equation}
V_\infty^4 = a_0 G M\; ,
\end{equation}
where $V_\infty$ is the asymptotic orbital velocity in the rotation curve of a galaxy and $M$ is its observable
mass. This equation is consistent with the Tully--Fisher relation between $V_\infty$ and the luminosity of a
galaxy, $L \propto V_\infty^\delta$, with $ 2.5 < \delta < 5$ \cite{TF}. The MOND theory gathered later some attention
\cite{bekenstein,KT,milgromc} but it is defended only by a limited number of theoreticians. The preference
for the dark matter explanation is not only a consequence of the application of Occam's razor principle. In recent
years the evidence for dark matter has been increasing by means of indirect observations that fit nicely into the
dark matter paradigm but which are very difficult to explain by the MOND theory: (i) Gravitational lensing by galaxy clusters
is far more intense than one could expect from the observable luminous matter \cite{tyson} (ii) The measurements of the
cosmic microwave background (CMB) anisotropies by the Boomerang and Wilkinson Microwave Anisotropy probes have placed certain
limits on the necessary amount of matter in the Universe to explain the formation of galaxies from the primordial fluctuations
at the matter--radiation decoupling era (see \cite{wmap3yr} for a three-year survey of the recent WMAP project 
and see \cite{boom} for a review of the Boomerang project). Moreover, the peak found at $l­_\mathrm{max} \simeq 200$ in the CMB spectrum implies
a flat Universe \cite{bernardis} and this can only be explained if we have more mass in the Universe than that observed in galaxies
and clusters.
Consequently, dark matter (DM) has become and essential ingredient of modern cosmology and without it concordance with the observational
data from all these independent sources cannot be achieved \cite{turner,liddle,coles}.
The overwhelming evidence for DM has promoted, from the last decade of past century, the proliferation of hypotheses on its nature. The
most economic hypothesis attributes this lost mass to bodies which emit little radiation: brown or white dwarfs, neutron stars
or black holes. However, the analysis of recent observations of gravitational microlensing events has discarded this source as a 
significant contribution to DM \cite{sumner,lesgourgues}.

A popular alternative assigns the main role to particles predicted by extensions of the Standard Model: axions (with a rest mass $\mu \sim 10^{-5}$ eV),
magnetic monopoles ($\mu \sim 10^{16}$ GeV), weakly interacting massive particles (WIMPs), the so-called neutralinos, or another supersymmetrical partner of
already known particles (with masses in the range $\mu \sim 1\mbox{--}10^3$ GeV) or, even, neutrinos heavier than ordinarily assumed ($\mu \sim 10$ eV).
This intriguing possibility has boosted an interesting synergy between cosmology, astrophysics and particle physics. In particular, there have been important
experimental efforts in the determination of an upper limit for the mass of the electron neutrino. In the Mainz neutrino
mass experiment \cite{mainz} and the Troitsk experiment \cite{troitsk} an absolute limit for the electron neutrino mass is found by investigating in large
detail the endpoint of the tritium $\beta$ decay spectrum. A limit $m­_{\nu,e} c^2 < 2.8 \mbox{ eV}$ is found in these experiments.
Even more precise measurements, in the sub-eV range are expected from Katrin project \cite{katrin}. Fixing confidence intervals for the masses of the
muon and tau neutrino is even more difficult as it has to be based on decay processes, such as $\tau \rightarrow 3\pi^{\pm}+\nu_\tau$, of
particles produced in accelerators \cite{cerutti}. This way it was estimated that $m_{\nu,\tau} c^2 < 15 \mbox{ MeV}$.
The phenomenom of neutrino oscillations also put some bounds on the splitting of masses for neutrinos \cite{2006review}. The maximum mass difference
squared, $\Delta m^2$ is bounded by $2.4 \times 10^{-3}\mbox{ eV}$. By combining this result with the upper limit on neutrino mass, a scenario in which
the three neutrino species have rest masses below the eV range is steadily gaining acceptance among cosmologists and particle physicists. Taking into
account that neutrino abundances are fixed in the Big Bang model, the known neutrinos cannot be the main ingredient of DM \cite{turner,liddle}. Experimental
settings devised to detect other candidates for DM particles -- axions or neutralinos -- have also been unsuccesful to date \cite{sumner}. The lack of
detection of WIMPs has favoured the suggestion of a family of sterile neutrinos -- sterile means that they do not interact via neutral or charged
currents -- which appears in some minimal extensions of the Standard Model \cite{asaka,shaposhnikov}. Sterile neutrinos are even more difficult to detect
than WIMPs because, essentialy, they only interact through gravity. Their masses are predicted to be in the range $2 < M_I < 5 \mbox{ keV}$. These masses, which
are relatively large compared with that of ordinary neutrinos, are acquired by means of the so-called seesaw mechanism \cite{chun}. It has been
proposed that sterile neutrinos could explain pulsar kick velocities \cite{kusenko} or the baryonic asymmetry in the Universe \cite{shaposhnikov}.

A pioneer proposal for a sterile neutrino was given by Dodelson and Widrow \cite{DW} in 1994. They considered that these neutrinos could be produced by
oscillations in the early Universe. However, a mass around $0.1$ keV was predicted in this model. Larger masses are desirable in order to confront the
main problem associated with light dark matter: the large free-streaming lengths which avoid the accretion of galaxies in the young Universe. Another 
mechanism was proposed later on by Shi and Fuller \cite{SF}. They proposed that sterile neutrinos with masses in a range $0.1$-$10$ keV are produced
via a lepton-number-driven resonant conversion of active neutrinos at the big bang nucleosynthesis epoch. With larger masses for the sterile neutrinos
the accretion problem could be circumvented. The energy spectrum of these neutrinos resembles a gaussian with a cutoff at $E/k_B T \simeq 0.7$.
Consequently, they are sufficiently cold to condensate on primordial fluctuations and favour galaxy formation. On the other
hand, the off-resonance process yields a different energy spectrum \cite{DW}:
\begin{equation}
f_{s}/f_{a}=\displaystyle\frac{7.7}{g_\star^{1/2}} \left(\displaystyle\frac{m_a}{1\mbox{ eV}}\right)^2 \left(\displaystyle\frac{1\mbox{ keV}}{m_s}\right) y 
\displaystyle\int_x^\infty \, \displaystyle\frac{d\xi}{(1+ y^2 \xi^2)^2}\; , 
\end{equation}
where $f_{s}$ ($f_{a}$) is the spectrum of the sterile (active) neutrinos, $m_s$ and $m_a$ are their masses, $g_\star$ is a constant, $y=E/(k_B T)$ and 
$x=78 (T/1 \mbox{ GeV})^3(1 \mbox{ keV}/m_s)$. For $T \ll 1 \mbox{ GeV}$ and a mass for the sterile neutrino of the order of keV the spectrum of sterile
neutrinos is proportional to that of the active species. Another alternative has been given by Shaposhnikov and Tkachev \cite{ST} and Kusenko \cite{HeavyScalar}.
In these models a heavy scalar decays through some channels into the sterile neutrinos. If this model is correct it would exist some chance for discovering
such a scalar in the Large Hadron Collider. More recently, it has been suggested that the decay of
sterile neutrinos into ordinary neutrinos and X rays will boost the production of molecular hydrogen in the early Universe \cite{BK}. Consequently, star formation will
also increase despite of the larger free-streaming lengths for Warm dark matter (WDM). In this scenario there is a constraint for X-ray observations which
implies that the mass of sterile neutrinos must be smaller than $3$ keV \cite{boyarsky}.

On the part of experimental particle physics, sterile neutrinos have been used in the sketching of an explanation  for the low energy anomaly in the neutrino
oscillations studied in the Liquid Scintillator Neutrino Detector or LSND \cite{gouvea}. Very recent results from the MiniBooNE collaboration excludes, at a
$98$ \% confidence, neutrino oscillations between two species as an explanation of the LSND anomaly \cite{miniboone}. The relevance of the sterile neutrino
theory for cosmology is also controversial \cite{dodelson}. 

On the other hand, some recent work by Boyarsky et al. \cite{lowerbound,lyman} determines the lower thresholds for the
masses of WDM particles based upon the Lyman-$\alpha$ forest and the WMAP5 results. This threshold depends on the production
mechanism of the sterile neutrino but a mass in the range of a few keV is compatible with the WMAP and Lyman-$\alpha$ data.

With independence of the nature of the constituent particles of DM, the DM hypothesis is the better explanation of a plethora of astrophysical observations.
However, an important point of discrepancy remains, because simulations of cold dark matter (CDM) accretion favour the formation of halos with a cusp at the 
center of the galaxies (the latest and more precise CDM's halos simulation is known as Via Lactea II \cite{kuhlen}). The calculations of galactic density profiles
from the rotation curves \cite{gilmore} and many observations of these rotation curves from galaxies \cite{blokbosma} contradict the prediction of an halo with
a peak at the center of galaxies. This phenomenon is related with Milgrom's law because a galactic core
almost free from DM will imply that there exists a critical orbit separating the inner regions of the galaxy, where rotation behaviour can be deduced from
the observed mass, and the outer regions where DM must be invoked to explain the discrepancy. The increasing and independent evidence on dark matter has made
the clarification of this paradox, in the context of dark matter models, urgent. Trying to reconcile DM models with Milgrom's law could also help us in clarifying
the properties of the fundamental particles which form it.

A recent proposal to explain Milgrom's law in a DM model has been given by Kaplinghat \& Turner \cite{KT}. These authors suggest that scale-free primordial
fluctuations and baryonic dissipation can explain the remarkable numerical coincidence $a_0 \sim c H_0$ between Milgrom's critical acceleration, $a_0$, and
the present value of Hubble constant, $H_0$. Dunkel has also shown that by taking into account the DM gravitation potential a generalized  MOND equation 
can be derived as a special limit \cite{dunkel}. However, the process which formed DM halos is not explained in these models.

We propose a different approach based upon a simple idea: warm dark matter (WDM) particles are trapped in the gravitational field of the galaxies
but their velocities are sufficiently large compared with the typical orbital velocities in the galactic cores and also random in their
directions. Consequently, they tend to evaporate from the core and distribute in a disperse halo. 
These WDM particles
were trapped back at the time of galaxy formation, when the CMB was roughly at a temperature $k_B T \simeq 3 \mbox{ eV}$ \cite{turner}. A kinetic model for the
Hubble expansion cooling of this collisionless gas indicates that the mass of these particles is in the keV range ($\mu \lesssim 3$ keV), in good agreement with
the sterile neutrino model. We use mean-field N-body simulations
for $2\times 10^4$ DM particles moving  in the static spherical potential well of the core in order to study the evolution of the halo and compare
with the mass distribution of our galaxy (Section \ref{Sect_II}).
A model with DM and cosmological constant fitting the most recent parameters obtained from WMAP \cite{wmap3yr} is used to study the
cooling of WDM in the early Universe.  The rest mass of the constituents of WDM is obtained by applying the condition that the typical velocity of the matter trapped
in primordial inhomogeneities was fixed at the galaxy formation era (Section \ref{Sect_III}). The reason for the quenching of the velocity distribution is the gravitational trapping of DM in
the galaxy. This way the effect of Hubble expansion after the galaxy formation era is eluded. The papers ends with some conclusions and
remarks in Section \ref{Sect_IV}.

\section{Self-gravitating model for Vlasov particles and a baryonic core}
\label{Sect_II}

\begin{figure}
\includegraphics{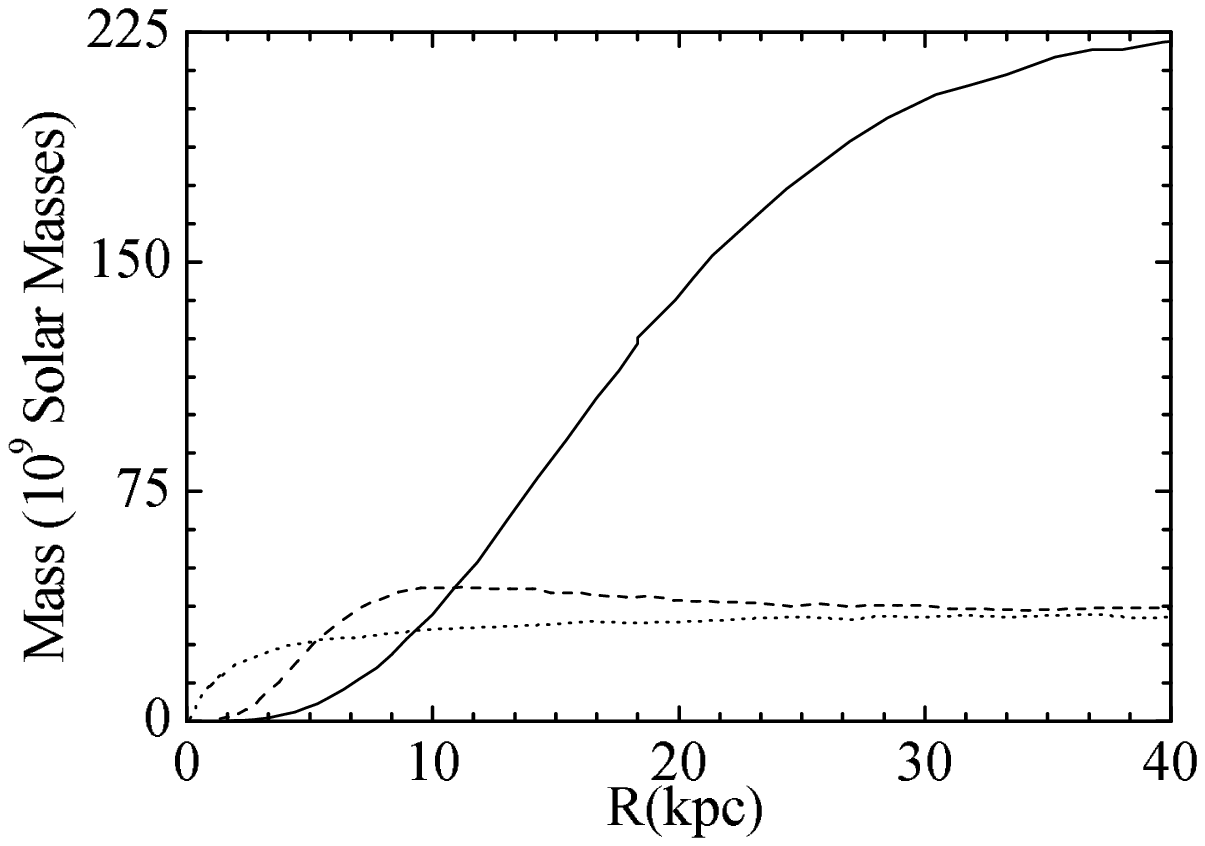}
\caption{Mass inside a sphere of radius $R$ centered at the center
of our Galaxy corresponding to the galactic bulge (dotted line), disk (dashed line) and
dark matter halo (solid line).\label{fig1}}
\end{figure}

Models for the matter distribution of galaxies have been developed
with a margin of accuracy only for the Milky Way. In this case the mass inside a sphere of radius $r$ from the galactic
center has been determined by Kalberla \& Kerp \cite{kalberla1,kalberla2}. We have plotted these results in Fig.~\ref{fig1} where separated curves
for the galactic bulge, the galactic disk and DM are shown. We notice that the DM mass inside an sphere of radius $< 5$ kpc is
negligible.

The mass distributions in Fig. \ref{fig1} have three conspicuous features: most of the visible mass is inside a sphere of radius $R_c \simeq 10 \mbox{ kpc}$ ($M_c=75 \times
10^9$ solar masses),
there exists a region corresponding to distances $r_0 \sim \mbox{ 5 kpc} < R_c$ almost free from DM and the halo spans from $r_0$ to very large distances
$R_0 \simeq 50\mbox{ kpc}$. The totalm mass of the galaxy (bulge, disk and halo) is, approximately, $M=300\times 10^9$ solar masses.

The objective of this section is to 
propose a dynamical model for the interaction of DM Vlasov particles and a spherical
baryonic core to explain qualitatively the DM distribution in Fig.\ \ref{fig1}. Self-gravitating systems have been an area of intense study for more than forty years \cite{aarseth,henon,lyndenbell,thirring}. Interest
into these models has been mainly spurred by their applications to globular clusters of stars \cite{meylanheggie}. Recently, the Smoluchowski-Poisson equation
for self-gravitating random walkers has been proposed as an adequate model for describing accretion of planetesimals in the solar nebulae where dissipation
and turbulence plays an important role \cite{chavanis,acedo}. In these works the qualitative behaviour of self-gravitating systems is derived from a
mean-field theory approach in which the gravitational force acting upon a particle is calculated by means of Gauss's theorem (for a spherically symmetric
system). This way we avoid a rigorous N body simulation which implies an unaffordable computational cost.

In order to develop our mean-field model for the evolution of the galactic halo we will take two assumptions for granted: (i) The galactic core is reasonably
well represented by an sphere with radius $R_c$ which contains most of the baryonic mass. This core was formed very early after the condensation around 
primordial inhomogeneities (ii) In the early Universe DM followed the distribution of baryonic matter more closely than today. In particular, we will assume
that all the DM was uniformly distributed inside the core. The velocity modulus, $\bar{v}$ is taken the same for all DM particles and the angular distribution
is homogeneous in the unit sphere. Alternatively, we can consider this initial condition as an extreme perturbation in the
configuration space without historical significance. In any case, we will find an evolution towards a fixed point in the
self-gravitation dynamics whose robustness is tested by exploring two initial velocity distributions a Dirac delta and 
a truncated parabola.

Starting from these initial conditions and, taking into account that DM particles are Vlasov particles that only interacts gravitationally among themselves
and with the core, the time evolution is deduced from the numerical integration of Newton equations by Euler method:
\begin{equation}
\label{Euler}
\begin{array}{rcl}
{\bf X}_i(t+h)&=&{\bf X}_i(t)+h {\bf V}_i(t)+{\cal O}(h^2) \\
\noalign{\smallskip}
{\bf V}_i(t+h)&=&{\bf V}_i(t)+h {\bf A}_i({\bf X}_1,\ldots,{\bf X}_n)+{\cal O}(h^2)\; , \; i=1,\ldots,N
\end{array}
\end{equation}
where ${\bf X}_i(t)$ and ${\bf V}_i(t)$ are, respectively, the position and velocity of the $i$-th particle at time $t$, $N$ is the total
number of particles, ${\bf A}_i({\bf X}_1,\ldots,{\bf X}_n)$ is the acceleration of the $i$-th particle due to the joint gravitational attraction
of the other Vlasov particles and the baryonic core and $h$ is the time step of the numerical method. It is convenient to scale these parameters in terms of 
characteristic parameters referred to 
the core. So, we will measure distances in units of $R_c$, masses in units of the mass of the core, $M_c$ and velocities in terms of the escape
velocity of a particle from the edge of the core in the absence of any more mass, $V_c=\sqrt{2 G M_c/R_c}$. The rest
of units are derived: $R_c/V_c$ is the unit of time and $V_c^2/R_c$ is our unit of acceleration. Iteration of Eq.\ (\ref{Euler}) is straightforward
if we assume that spherical symmetry is preserved by the evolution (the acceleration for each particle always points towards the center of the core). This way, we
can easily calculate a mean-field estimation of the modulus of ${\bf A}_i$ as follows:
\begin{equation}
\label{AVlasov}
\left\vert {\bf A}_i \right\vert=\left\{\begin{array}{rcl}
\displaystyle\frac{r_i}{2}+\displaystyle\frac{n(r_i) m}{2 r_i^2} \quad &\mbox{if}&\; r_i < 1\\
\noalign{\smallskip}
\displaystyle\frac{1+n(r_i) m}{2 r_i^2} \quad &\mbox{if}&\; r_i \ge 1\; ,
\end{array}\right.
\end{equation}
where $n(r)$ is the number of particles at a distance from the center of the core smaller than $r$, $r_i$ is the distance of the $i$-th particle
from the center of the core at time $t$. $m=M_H/N$ is the scaled mass of a single particle in the simulation, $M_H$ being the mass of the DM halo. We must take into
account that these particles would represent many real DM particles because $N$ is small.  Notice than in Eq.\ (\ref{AVlasov}) we have already used the scaled 
parameters defined above. We will use this units in the following.

\begin{figure}
\includegraphics{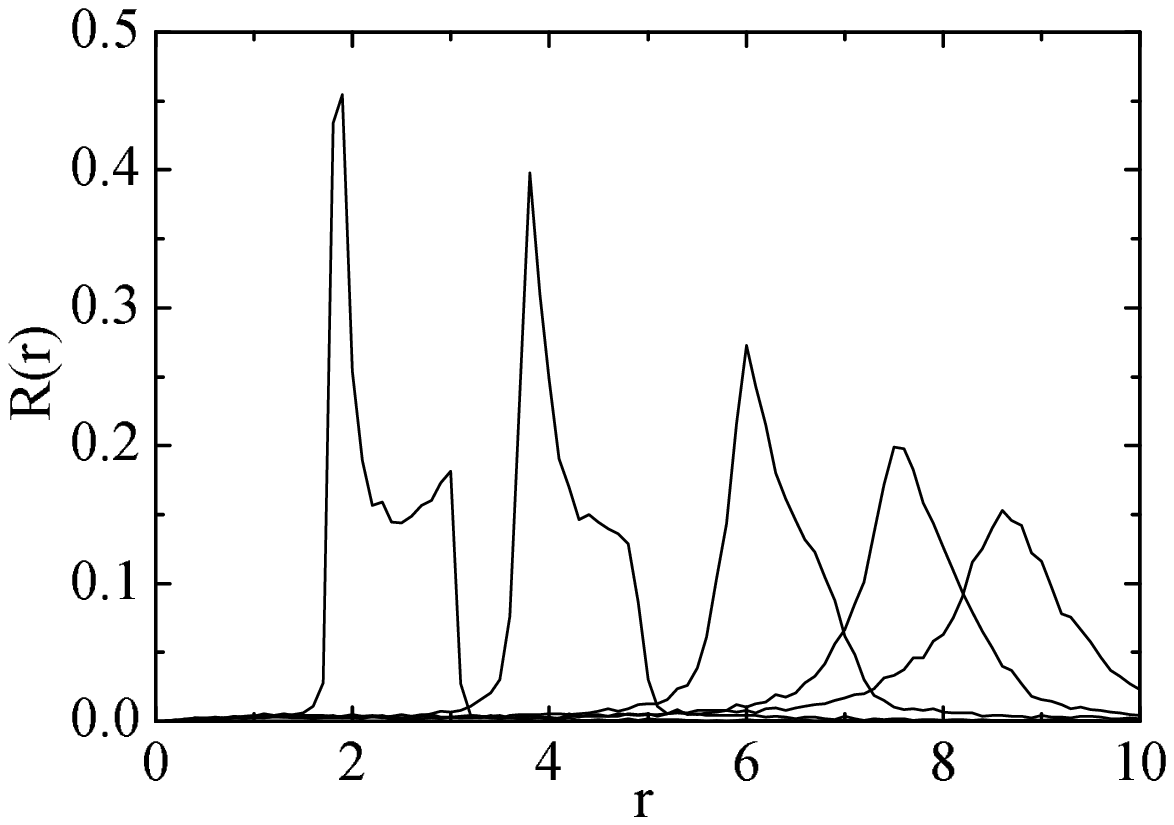}
\caption{The radial density of DM particles for $t=2$, $5$, $10$, $15$, $20$ (numbered from the left to right according to
the position of the peaks). We take $\bar{V}=1.5$ as the initial velocity modulus.\label{fig2}}
\end{figure}

In order to grasp the behaviour of the halo for large times we have performed simulations for several values of the typical velocity of DM particles and
$N=20000$. Timestep is taken as $h=10^{-4}$. The results for the evolution radial density as a function of time for $\bar{V}=1.5$ are shown in
Fig.\ \ref{fig2}. The radial density, $R(r)=4 \pi r^2 \rho(r)$ ($\rho(r)$ being the volumetric density), was calculated using the following approximation
$R(r) \approx (n(r+\Delta r)-n(r)) m /\Delta r$ and taking $\Delta r=0.1$. We find that for $\bar{V}=1.5$ and larger typical velocities the DM evaporates from
the baryonic core. In this process a diffusive wave with increasing width develops. This spherical wave propagates at almost constant velocity expanding forever
into intergalactic space. Globular clusters and dwarf galaxies should have lost most of their DM content through this evaporation process. 
However, dwarf galaxies are also DM dominated. The mechanism in this case could be peculiar and related with the evolution
of this type of galaxies as we discuss in Sec. \ref{Sect_IV}. If we apply the result of this simulation
to the case of $\omega$ Centauri using the data given by Meylan and Mayor \cite{meylan}, $R_c=30$ pc and $M_c=18$ millions solar masses, we find that the process in 
Fig.\ \ref{fig2} happened in only $8$ million years. In the course of this time the halo has expanded to a scale of $\simeq 300$ pc. Later on, the halo continued its
expansion being finally captured by the potential well of the Milky Way or dispersed into the intergalactic space of the Local Group. This is a similar process
to the formation of nebulaes, such as the Crab or the Ring nebulaes, by the explosion of supernovaes \cite{hanka}. In this processes the debris of the supernovae
explosion, moving at a speed larger than the escape velocity from the original star, disperse in a diffuse halo over the centuries to form the nebulaes we see
today.

\begin{figure}
\includegraphics{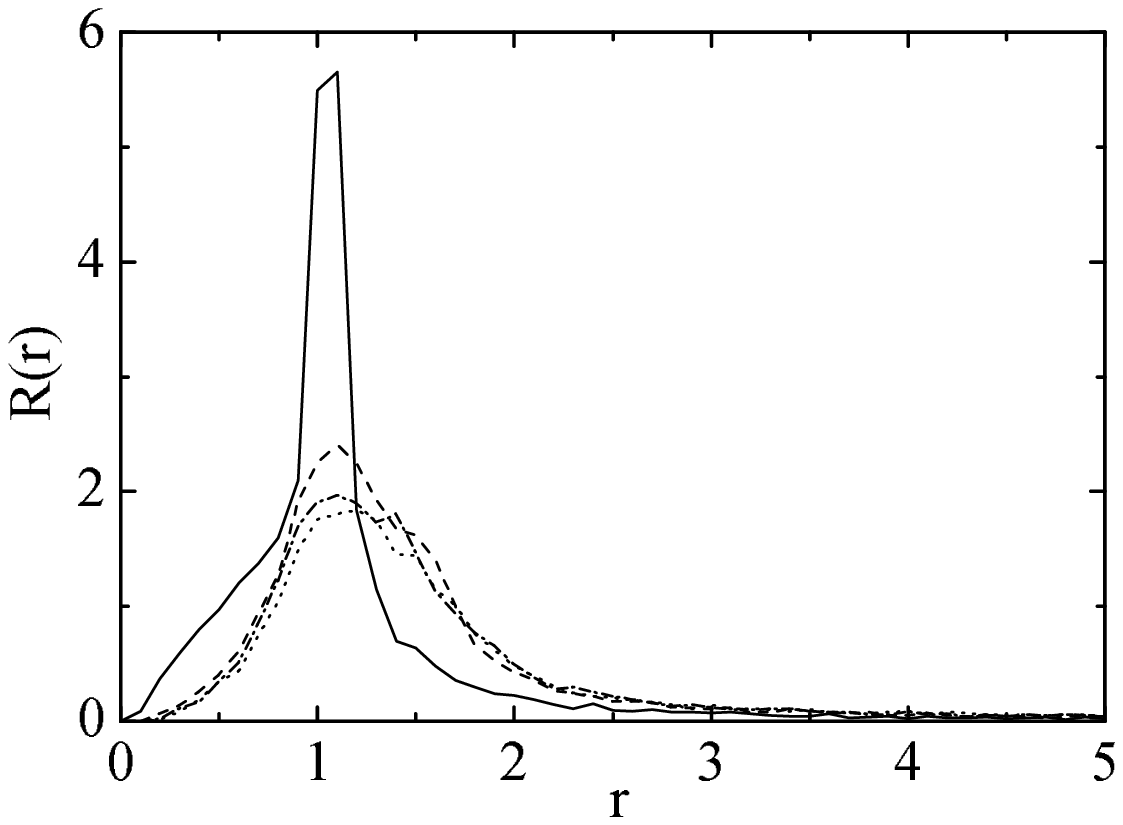}
\caption{The radial density of DM particles for $t=5$ (solid line), $20$ (dashed line), $35$ (dashed-dotted line) and $50$ (dotted line).  Initial velocity modulus
was $\bar{V}=1.1$.\label{fig3}}
\end{figure}

On the other hand, galaxies have conserved a great amount of DM. This behaviour is obtained for smaller typical velocities for the DM particles. In Fig.\ \ref{fig3}
we have plotted the results for the radial density of the DM distribution in four time snapshots from $t=5$ to $t=50$ with $\bar{V}=1.1$. The development of a characteristic cusp
structure, with a cusp at a finite distance from the galactic center, is clearly observed. Moreover, the Vlasov particles do not disperse but, equally important, they do not collapse at the center of the core. An stationary
state is achieved with a maximum radial density at the edge of the baryonic core. The resulting DM halo keeps bound by the joint attraction of the core and its own
self-gravitation. In the case of the Milky Way we may assume $R_c=10$ kpc and $M_c=75\times 10^9$ solar masses (using this radius most of the baryonic mass is contained 
inside the core). The time evolution of the simulations in Fig.\ (\ref{fig3}) should correspond to, approximately, 2000 millions years.

\begin{figure}
\includegraphics{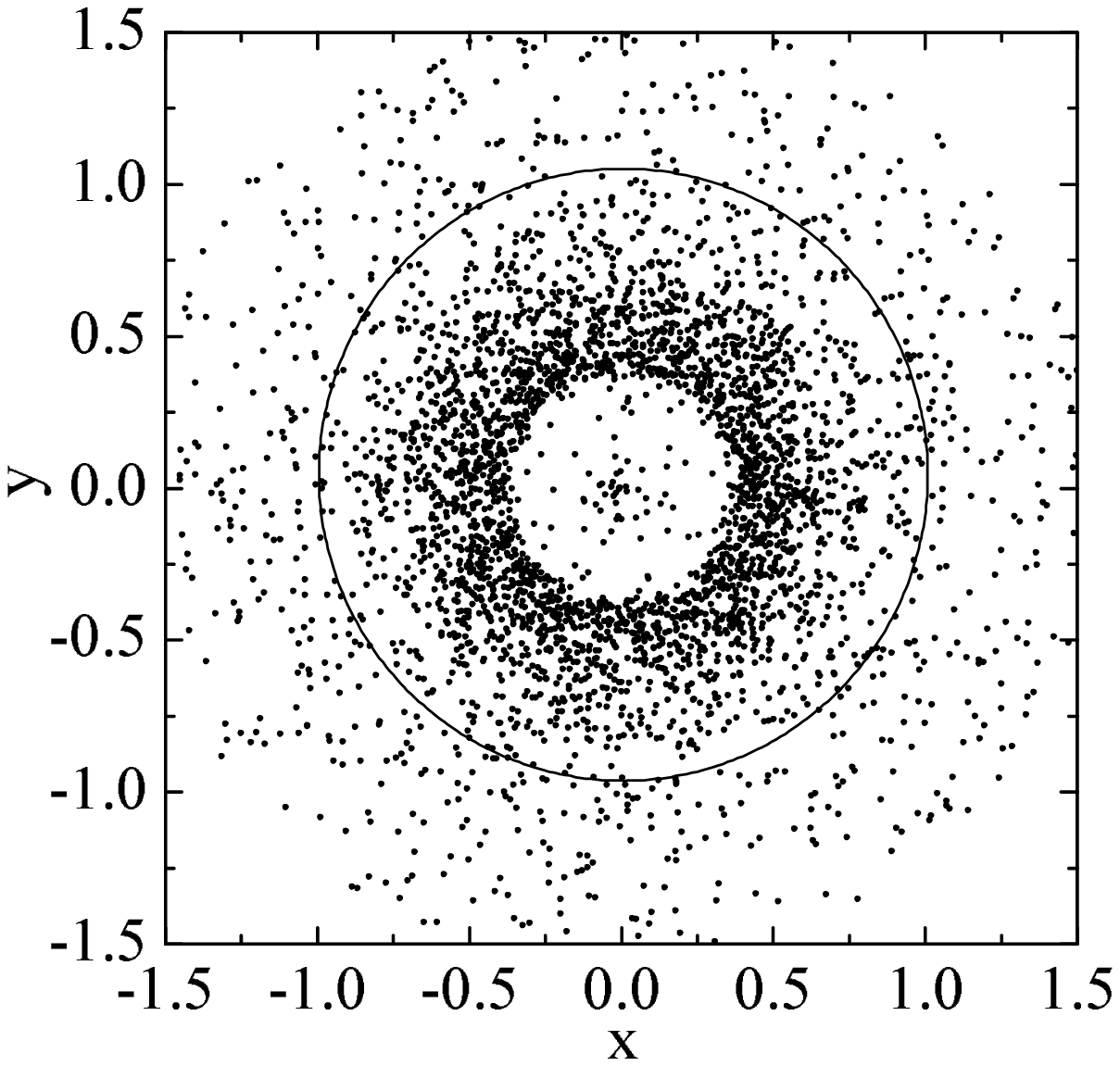}
\caption{Particles in the DM halo at $t=10$ with a coordinate $-0.2 < z < 0.2$ (initial velocity $\bar{V}=1$). The circle delimits the extension of the baryonic
core.\label{fig4}}
\end{figure}

An slice of the halo at $t=10$ obtained by plotting only the particles with coordinate $z$ in the range $[-0.2,0.2]$ is shown in Fig.\ \ref{fig4}. We observe the
development of an inner spherical region with it is already almost devoid of DM particles.

In order to compare with the Kalberla \& Kerp density model for out Galaxy \cite{kalberla1,kalberla2} we have fitted their data by a Pad\'e approximant as follows:
\begin{equation}
\label{pade}
{\cal M}_H(r)=\displaystyle\frac{M_H r^3}{\xi_0+\xi_1 r+\xi_2 r^2+r^3} \; ,
\end{equation}
where ${\cal M}_H(r)$ is the DM mass inside a sphere of radius $r$,  ${\cal M}_H$ is the total mass of the halo and $\xi_0$, $\xi_1$ and $\xi_2$ are constants. If we
measure the mass in units of $10^{12}$ solar masses and the distance in kpc we get ${\cal M}_H=0.225$, $\xi_0=5874.07$, $\xi_1=-40.17$ and $\xi_2=-0.2539$. These 
values were obtained by imposing the conditions that the result of Eq.\ (\ref{pade}) coincides with the data in Fig.\ \ref{fig1} for $r=5$, $10$ and $20$ kpc. The mass
of the halo is $M_H=M-M_c=225 \times 10^9$ solar masses.

\begin{figure}[pb]
\includegraphics{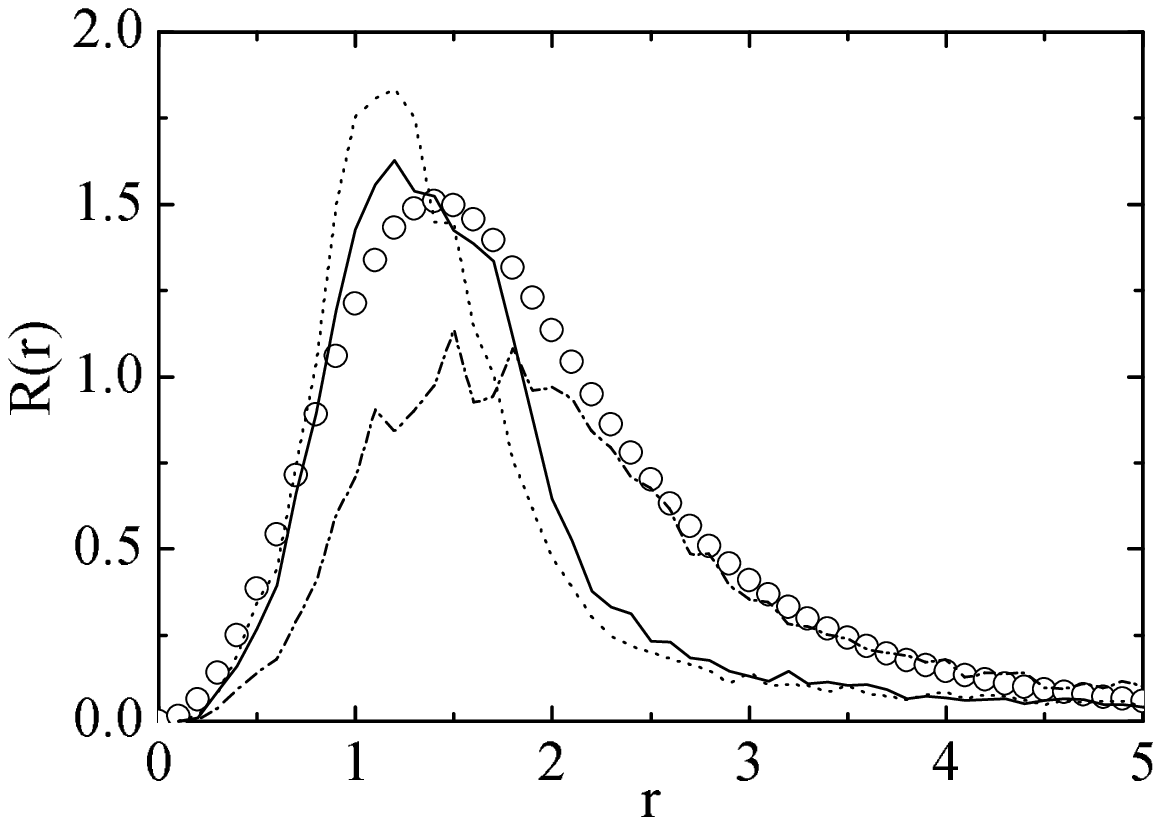}
\caption{The asymptotic form of the galactic radial density of the DM halo for $\bar{V}=1$ (dotted line), $\bar{V}=1.15$ (solid line) and $\bar{V}=1.3$ (dashed-dotted line).
Circles correspond to the Pad\'e approximant for the density of the Milky Way DM halo as derived from Eq. (\protect\ref{pade}). Initial DM mass is three
times larger than the baryonic mass and we use $10^4$ particles. The radial distance $r$ is measured in units
of $R_c=10$ Kpc.\label{fig5}}
\end{figure}

In Fig.\ \ref{fig5} we compare the results of the Pad\'e approximant for the radial density, $R(r)=d {\cal M}_H(r)/d r$ with the simulation results for three different
initial velocities of the DM particles (distances and masses are conveniently scaled with $R_c=10$ kpc and $M_c=75\times 10^9$ solar masses) after $500000$ time steps
($t=50$). The initial mass of the DM was $M_H=3$ (in units of the mass of the baryonic core). The position and radial density at the cusp are qualitatively well described
by the distribution obtained in the simulations for $\bar{V}=1.15$. However, the tail
of the halo is probably composed by particles with an initially larger velocity and cannot be fitted by considering a Dirac delta velocity distribution. However, we obtain
an  estimation of $\bar{V}=1.3 V_c=330$ km s$^{-1}$ for the primordial typical velocities of DM particles.

We also notice that the area under the predicted profiles in Fig.\ \ref{fig5} is smaller than the initial mass of the DM. The reason for that is the evaporation
to intergalactic space of a $20$ \% of the DM particles after the virialization of the halo. 

It is also interesting to consider a more realistic initial velocity distribution for the DM particles. In this spirit, we propose a distribution proportional
to the square of the velocity modulus with a cut-off as follows:
\begin{equation}
\label{vdist}
f(v)=\left\{ \begin{array}{rcl}
\displaystyle\frac{81}{64} \displaystyle\frac{v^2}{\bar{v}^3} &\quad& \mbox{ if $v < 4 \bar{v}/3$} \\
\noalign{\smallskip}
0 &\quad& \mbox{ if $v > 4 \bar{v}/3$}\; ,
\end{array}\right.
\end{equation}
The cut-off is chosen to verify the normalization condition and $\bar{v}$ is the average velocity. Sterile neutrinos produced by the resonant conversion of active
neutrinos are supposed to end up with an energy spectrum of this kind in the protogalaxies \cite{SF}.

\begin{figure}
\includegraphics{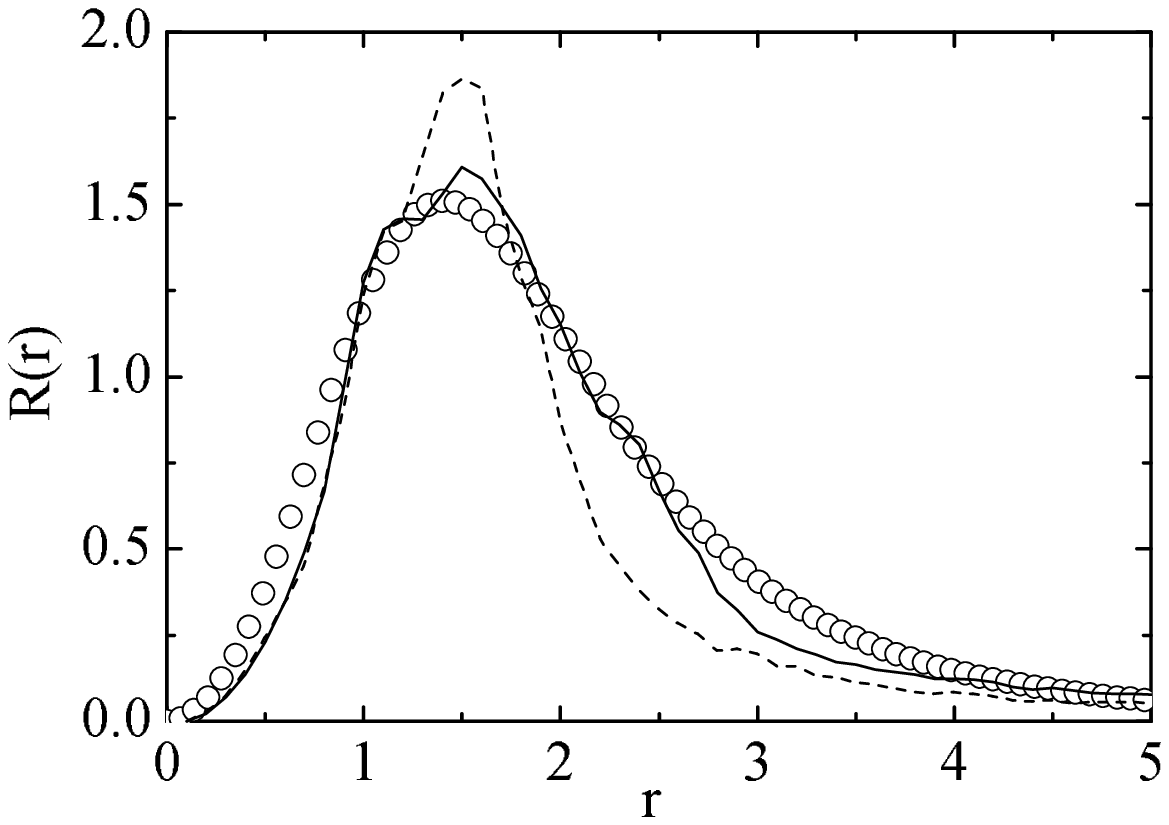}
\caption{The same as Fig. \protect\ref{fig5} but for an initial distribution of velocities as given in Eq. \protect\ref{vdist}. The dotted line corresponds
to $\bar{v}=1.2$ and $M_H=3$ and the solid line corresponds to $\bar{v}=1.3$ and $M_H=4$. We used $500000$ time steps to achieve the stationary state and 
$10^5$ DM particles. The unit of distance is $R_c=10$ Kpc.\label{fig6}}
\end{figure}

In Fig.\ \ref{fig6} we have plotted the simulation results for an initial homogeneous distribution of $10^5$ DM particles moving with initial velocities drawn
according to Eq.\ (\ref{vdist}). The agreement with the radial density of our galaxy is specially good if we consider a primordial typical velocity $\bar{V}=1.3 V_c$ 
a mass of the DM halo four times larger than the mass of the baryonic matter.

In the next section we deduce the mass of DM particles from this typical velocity by considering
that they cooled by the expansion of the Universe before being captured by the protogalaxies.

\section{Hubble cooling of dark matter in the early Universe}
\label{Sect_III}
In this section we discuss a kinetic model for weakly interacting massive particles or free Vlasov particles only interacting gravitationally as they
move in a $\Lambda$CDM expanding Universe. We assume a flat Universe filled with the critical mass according to accepted concordance with the CMB
observations \cite{wmap3yr,bernardis}. The space-time metric is given by: 
\begin{equation}
d s^2=d \tau^2-a^2(\tau)\left[ d x^2+d y^2+d z^2\right] \; .
\end{equation}
A standard result of cosmological models \cite{weinberg,gravitation} is the following set of equations for the cosmological radius, $a(\tau)$, in terms of the energy
density, $\epsilon$, and the pressure, $p$:
\begin{eqnarray}
\label{aevol}
\displaystyle\frac{\ddot{a}}{a}&=&-\displaystyle\frac{4 \pi G}{3 c^4} \left(\epsilon+3 p\right) \\
\noalign{\smallskip}
\left(\displaystyle\frac{\dot{a}}{a}\right)^2&=&\displaystyle\frac{8 \pi G}{3 c^4} \epsilon\; ,
\end{eqnarray}
where the dots denote successive derivatives with respect to $\tau$. The parameter $H=\dot{a}/a$ is
the so-called Hubble constant whose value has been largely constrained by the WMAP project \cite{wmap3yr} and is currently
accepted to be given by $H=73$ km s$^{-1}$ Mpc$^{-1}$. From Eq.\ (\ref{aevol}) we find that the energy density of the Universe
is related with $H$ as $\epsilon=3 c^4 H^2/ 8 \pi G$. In the $\Lambda$CDM model, the energy content of the Universe comes essentially,
from dark energy (with an equation of state $p=-\epsilon$) and dark matter or baryonic matter (whose pressure is negligible compared
with the energy density and is, usually, ignored, $p=0$). In percentage terms, WMAP results are compatible with the values: $\Omega_\Lambda=0.72$,
$\Omega_{\mathrm{DM}}=0.23$ and $\Omega_{\mathrm{B}}=0.05$ for the fractions of dark energy, dark matter and baryonic matter, respectively.

Integration of the system in Eq.\ (\ref{aevol}) yields:
\begin{equation}
\label{astar}
a^\star(\tau)=\left(\displaystyle\frac{1-\Omega_\Lambda}{\Omega_\Lambda}\right)^{1/3}\, \sinh^{2/3} \left[ \displaystyle\frac{3 \tau}{2\tau_\Lambda}\right] \; , 
\end{equation}
where $a^\star(\tau)$ is the cosmological radius at cosmological time $\tau$ measured in units of the present radius and $\tau_\Lambda=1/(H \sqrt{\Omega_\Lambda})\simeq
15.8\times 10^9$ years. From the condition $a^\star(\tau_0)=1$ we get the age of the Universe, $\tau_0\simeq 13.2\times 10^9$ years.
The absolute curvature radius of the Universe at present, $a(\tau_0)$, can be inferred from the decceleration parameter \cite{riess,perlmutter1} but the value of $a(\tau_0)$ is not
necessary in our kinetic model.

Standard Big Bang Cosmology placed the era of formation of galaxies at the time corresponding to a temperature of the CMB $k_{\mathrm{B}} T \approx 3$ eV. At that time, matter energy density
exceeded radiation energy density and collapse around primordial fluctuations started \cite{turner}. The CMB temperature is governed by a red-shift law: $T(\tau)=2.7277/a^\star\, \mathrm{K}$  which
in connection with Eq. (\ref{astar}) implies that the protogalaxies formed $12000$ years after the Big Bang. If dark matter were composed by weakly interacting massive particles
it is also generally accepted by cosmologists that decoupling took place at a CMB temperature $k_{\mathrm{B}} T \approx 1$ MeV, i.e. $30$ minutes after the Big Bang.

In our kinetic model, we consider that a DM particle colliding with a baryon (interacting via weak fields) at the time before galaxy formation thermalizes and, consequently,
its kinetic energy raises to $k_{\mathrm{B}} 2.7277/a^\star$ eV, because baryons and radiation are in thermal equilibrium. The total energy of a DM particle after an interaction 
is taken to be
\begin{equation}
\label{Etot}
E=\mu c^2+\displaystyle\frac{k_\mathrm{B} T}{a^\star}\; ,
\end{equation}
where $\mu$ is the rest mass of the DM particle. Linear momentum at that time is given by
\begin{equation}
\label{momentum}
p^2=\displaystyle\frac{k_\mathrm{B} T}{a^\star}\left[2 \mu+\displaystyle\frac{k_\mathrm{B} T}{a^\star c^2}\right] \;.
\end{equation}
Linear momemtum of a free particle in an expanding Universe decreases as $a^{-1}(\tau)$  \cite{carroll}. Consequently, the kinetic energy of a DM particle at time
$t$, that thermalized at time $t'$ and never collided again, can be expressed as follows
\begin{equation}
\label{Qttp}
Q(t,t')=\sqrt{\mu^2 c^4+\left[\displaystyle\frac{a^\star(t')}{a^\star(t)} \right]^2 \epsilon_{\mbox{rad}} (t') \left[\epsilon_{\mbox{rad}}(t')+2 \mu c^2\right]}-\mu c^2\; ,
\end{equation}
where $\epsilon_{\mbox{rad}}(t)=\epsilon_0/a^\star(t)$ and $\epsilon_0=2.35046 \times 10^{-4}$ eV is the average energy of a CMB photon in present day Planck spectrum. The
average kinetic energy of a DM particle at time $t$, assuming that it was thermalized at an earlier time $t'$, is approximately given by
\begin{equation}
\label{avK}
\left\langle K(t) \right\rangle=e^{-\left\langle \lambda \right\rangle (t,t') (t-t')} Q(t,t')+
\displaystyle\int_{t'}^t \, d \eta \lambda(\eta) e^{-\left\langle \lambda \right\rangle (\eta,t) (t-\eta)} Q(t,\eta)\; ,
\end{equation}
where $\left\langle \lambda \right\rangle(t,t')$ is the average collision frequency in the time interval $(t',t)$ for a DM particle in the baryonic background. Notice that
the first term gives us the contribution for those DM particles that do not collide in that interval of time and the second corresponds to DM particles colliding one of
several times with baryons (with the most recent collision at time $\eta$).

Collision frequency for weakly interacting dark matter particles in a baryonic background is estimated as follows:
\begin{equation}
\label{collfreq}
\lambda(t)=\sigma(E) n_B(t) v_B(t)\; ,
\end{equation}
where $\sigma(E)$ is the cross section for weak interactions \cite{liddle}:
\begin{equation}
\label{sigmaweak}
\sigma(E)\approx (\hbar c)^{-4} G_F^2 E^2\; .
\end{equation}
For sterile neutrinos weak interactions are suppressed by a factor $\sin^2 \theta$, $\theta$ being the mixing angle with
ordinary neutrinos \cite{BK}. In the oscillation production theories an upper limit has been derived in order to
avoid the crowding of the early Universe with sterile neutrinos \cite{abazajian}: $\theta < 1.3 \times 10^{-4}(1\mbox{ keV}/ M_s)^{0.8}$, 
where $M_s$ is the mass of the sterile neutrino.
In the early Universe the energy, $E$, is commonly identified with the
CMB background enery scale, $E=\epsilon_0/a^\star(t)$. $G_F=1.166 \times 10^{-5} (\hbar c)^3$  GeV$^{-2}$ is Fermi's constant. The baryonic density
decreases with the cube of the Universe radius:
\begin{equation}
\label{nB}
n_B(t)=\displaystyle\frac{n_0}{a^{\star 3}(t)}\; ,
\end{equation}
using a present day baryon density, $n_0\simeq 3$ baryons m$^{-3}$ as 
readily deduced from the abundance $\Omega_B \simeq 0.05$ for baryons at the
$\Lambda$CDM model. Finally, $v_B(t)$ is the relative velocity between DM
particles and baryons. If we take it as the baryon velocity (a rough approximation)
we get
\begin{equation}
\label{vB}
v_B(t)=c \displaystyle\frac{\sqrt{\epsilon_0(t)(\epsilon_0(t)+2 m_B c^2)}}{\epsilon_0(t)+m_B c^2} \approx \sqrt{\displaystyle\frac{2 \epsilon_0}{m_B c^2}} a^{\star-1/2}\; ,
\end{equation}
where $m_B c^2 \simeq 939$ MeV is the baryon mass. Inserting Eqs. (\ref{vB}), (\ref{nB}) and (\ref{sigmaweak}) into the expression for the collision frequency
in Eq. (\ref{collfreq}) yields 
\begin{equation}
\label{lambdat}
\lambda(t)=(G_F \epsilon_0)^2 (\hbar c)^{-4} n_0 \sqrt{\displaystyle\frac{2 \epsilon_0}{m_B}} a^{\star -11/2}\; .
\end{equation}
From Eqs.\ (\ref{lambdat}) and (\ref{astar}) we can obtain the time average collision frequency as follows
\begin{equation}
\langle \lambda \rangle(t,t')=\displaystyle\frac{1}{t-t'}\, \displaystyle\int_{t'}^t \, \lambda(t) d t\; .
\end{equation}
This integral can be evaluated exactly in terms of hypergeometric functions yielding
\begin{equation}
\label{lambdavg}
\begin{array}{rcl}
\langle \lambda \rangle (t,t')&=&\displaystyle\frac{2}{3} \displaystyle\frac{\tau_\Lambda}{t-t'} \left( G_F \epsilon_0\right)^2 (\hbar c)^{-4} n_0 
\sqrt{\displaystyle\frac{2 \epsilon_0}{m_B}} \\
\noalign{\smallskip}
& &\left( \displaystyle\frac{\Omega_\Lambda}{1-\Omega_\Lambda} \right)^{11/6} \left[
\psi\left( \displaystyle\frac{3 t}{2 \tau_\Lambda}\right)-\psi\left( \displaystyle\frac{3 t'}{2 \tau_\Lambda}\right)\right] \; ,
\end{array}
\end{equation}
where
\begin{equation}
\label{psix}
\begin{array}{rcl}
\psi(x)&=&\displaystyle\frac{1}{16} \cosh x \left[ 5 e^{-2 \pi i/3} {}_2F_1\left(1/2,1/3,3/2,\cosh^2 x\right) \right. \\
\noalign{\smallskip}
 & &+\left. 3 \displaystyle\frac{5 \sinh^2 x-2}{\sinh^{8/3} x}\right]\; ,
\end{array}
\end{equation}
${}_2F_1$ being an hypergeometric function of order $(2,1)$ \cite{abramo}.
\begin{figure}
\includegraphics{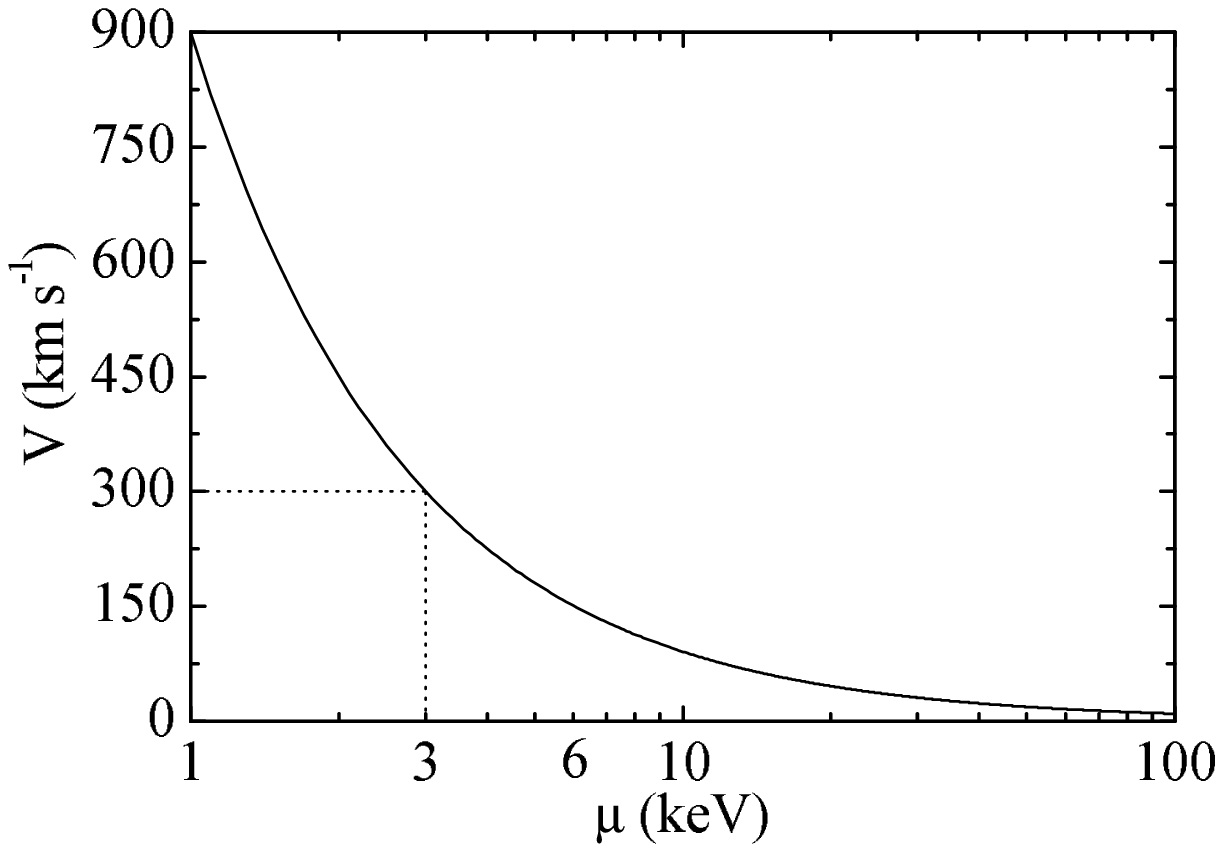}
\caption{Typical velocity of DM particles at the galaxies formation era as a function of its rest mass. Solid line is the result for sterile
neutrinos (which we assume to decouple at a temperature correspondig to $150$ MeV).\label{fig7}}
\end{figure}

In Fig. \ref{fig6}, we have plotted the typical velocity of DM particles as a function of its rest mass deduced from its kinetic energy in Eq. (\ref{avK}):
$v/c=\sqrt{\left\langle K(t) \right\rangle^2+2 \left\langle K(t) \right\rangle \mu c^2}/(\left\langle K(t) \right\rangle+\mu c^2)$. The initial time is approximately
$t_i=1$ second after the Big Bang (corresponding to $k_B T \simeq 150$ MeV, the estimation for the decoupling era of sterile neutrinos produced
via oscillations \cite{DW,SF}). Notice that the average kinetic energy will be given by Eq.\ (\ref{Qttp}) because sterile neutrinos are 
collisionless Vlasov particles after their decoupling. Collision frequency is very small indeed: if we consider that the mixing
angle for sterile neutrinos is $\theta \simeq 10^{-4}$, according to a recent estimation for active-sterile neutrino oscillations \cite{abazajian,ASK}, the average
collision frequency in the interval $t_i < t < 1.1 t_i$, calculated from Eq.\ (\ref{lambdavg}), is of the order of a collision every $10^7$ years.
In our scenario, the primordial typical velocities must be $\simeq 300$ km s$^{-1}$  which implies, 
according to Fig.\ (\ref{fig7}),  that their rest mass is roughly $3$ keV, in good agreement with the estimation for the mass of sterile neutrinos.
On the other hand, the results in Fig.\ \ref{fig7} are very robust concerning the era of decoupling of sterile neutrinos or their mixing angle (we conclude that sterile
neutrinos were free streaming particles almost from their appearance). Consequently, the decision about the scenario most suitable for the formation of sterile neutrino:
off-resonance\cite{DW}, on-resonance\cite{SF} or by the decay of heavy scalars \cite{ST,HeavyScalar} is to be given by particle phyics. On the other hand, a recent lower
bound $m_{\mbox{RP}}=2$ keV for the resonance production mechanism and $m_{\mbox{NPR}}=1.77$ keV for the non-resonant mixing with active neutrinos obtained from a phase-space
analysis of DM distribution in dwarf spheroidal galaxies \cite{lowerbound} agrees with our estimate from the proposed dynamical self-gravitating model.

\section{Conclusions and Remarks}
\label{Sect_IV}

We propose that dark matter trapped in galaxies is warm and evaporates from the galactic cores.
The plausibility of this scenario is investigated in the context of a self-gravitation 
model for the halo of DM particles moving in the potential well of a spherical baryonic core.

The distribution of DM in galaxies is inferred from their rotation curves \cite{faber,rubin,caldwell}. There is some
agreement about that DM does not concentrate in the core of the galaxies, as one could naively expect, but their mass
density peaks at a finite distance from the galaxies center -- varying from a galaxy to another -- and few
DM particles are found inside the galactic core. This is confirmed by recent observations of $26$ low surface brightness galaxies
carried out by de Blok \& Bosma \cite{blokbosma} and by calculations of galactic density profiles inferred from rotation 
curves \cite{gilmore}. This is the so-called cuspy halo problem.
Nevertheless, there is also a controversy about the interpretation of galactic rotation curves and some
authors claim that central rotation velocities are underestimated by conventional techniques \cite{rhee,valenzuela}. 

In our theory, Milgroms' law and the cuspy halo problem are related. We show that it can be explained within the dark matter paradigm by
simply assigning a primordial typical velocity to dark matter particles within the order of $300$ km s$^{-1}$. The self-gravitation model predicts the evolution 
towards a cuspy halo (with a cusp at a finite distance from the galactic center) starting from an initial homogeneous distribution. In this process an inner spherical region, with a radius similar to the radius of the core, appears. This region
is almost empty of DM particles and, consequently, anomalies in the orbits of visible matter are only detected at a critical radius corresponding to Milgrom's orbital acceleration.
Assuming that the velocity distribution of the dark matter gas was quenched at the time of the formation of galaxies, we have derived
the typical mass of dark matter particles to fulfill the condition above. Their mass ($\sim 3$ keV) coincides with recent estimations for sterile neutrinos which arise
in extensions of the standard model of elementary particles.
In order to develop a better insight into the evolution of galaxies some simplifying assumptions used in our model could be removed in more realistic implementations, i. e.,
we can consider a rotating disk of baryonic matter in addition to the core, the parallel evolution of the halo of Vlasov particles and the dissipative clouds of
baryons. Another problem is the dominance of DM over baryonic matter in dwarf galaxies with typical velocities in the
flat region of rotation curves $\simeq 60$ km s$^{-1}$ \cite{blokbosma}. An special mechanism could be considered in this
case such as a WDM$+$CDM model or the accretion of low velocity DM particles in the outskirts of large galaxies nearby.
With these features a better understanding on
the formation of galaxies and, in turn, on the nature of DM could be obtained. Work along these lines is in progress and will be published elsewhere.

\subsection{acknowledgments}

The author gratefully acknowledges M. Tung for some useful suggestions and a critical reading of
the manuscript. The NASA's Astrophysics Data System is also acknowledged for providing some references.

\end{document}